\documentstyle[12pt]{article}

\begin{document}
\title{Predicting and characterizing data sequences from structure-variable
 systems}
\date{}
\author{H. P. Fang$^{a,b,*}$  and L. Y. Cao$^b$
\\{\small  $^a$CCAST (World Laboratory) P. O. Box. 8730,
	Beijing, 100080, China;}
	\\{\small $^b$Institute of Theoretical Physics, P.O. Box. 2735,
	Beijing, 100080, China.}
	}
\maketitle
\begin{abstract}
In principle,  all the  natural systems such as  biological, ecological
 and economical systems are structure-variable systems (in which some
environment parameters  are not fixed). In this Letter we show that
data sequences  from many structure-variable systems are short-term
predictable. We also argue   how to  characterize the   data sequences
 from structure-variable systems.
\end{abstract}
\vskip.15in


In the last two decades, there has been rapid progress in understanding
 deterministic chaos, not only in theoretical modelling [1], but also in
  experimental testing [2]. Since the the pioneering paper in [3]
  and an embedding theorem due to Takens [4], it becomes
 widely known that the current state of the finite-dimensional dynamical
  systems can be identified using a vector of time series measurements.
In the proxy state space consisting of the delay-coordinate vector,
analysis of the topological properties of the chaotic
 attractor underlying the time series is  performed. The application of
 this idea includes noise filtering [5],  control of unstable periodic
orbits solely from a time series record [6], and prediction of the chaotic
time series[7-8].

Up to now, all the studies and the applications are restricted to the
chaotic dynamics of structure-invariable systems (SIS)  (in which all the
 parameters are fixed)  or assumed structure-invariable systems [9-11].
  However, many natural systems such as  biological, ecological
and economical systems belong to structure-variable systems (SVS (in which
some parameters are not fixed), they have developed into other ones before
they reach equilibrium states. The stock system, the financial
expenditure system for one country, the seismic system and the climatic
 system are examples since the environments of these systems
changes rather rapidly before they settle down at some asymptotic states.
Remarkably, it was found recently [12] that the Chinese national financial
expenditure from 1973 to 1992  can be well predicted based on the records
in 1953-1972 by a nonlinear prediction
algorithm [8]. It is well known that the Chinese economical environment
had been changed greatly from 1953 to 1992. One may ask a question.
Why does a SIS prediction algorithm work so well? The first motivation
of this Letter is to present an approach to answer this question and then the
numerical results in [12] can be understood. We will use the  concept of
developing diagram [13] to illustrate our idea. It is  found that every SVS
can be transferred into a corresponding higher-dimensional SIS by adding some
new variables so that the nonlinear SIS prediction algorithms can work
well for
SVSs.  Moreover, in the last decade, there has been a lot of work
 trying to apply the chaotic dynamics to many natural systems [e.g.,9-11].
The correlation dimensions ($D_2$) and the largest Lyapunov exponents
($\lambda$) are calculated with the familiar algorithms
for chaotic attractors in SISs, and the existence of those quantities has
been considered as evidence of chaotic dynamics in those natural systems.
In principle, many of these natural systems are SVSs. Only when the environment
 parameters change sufficiently slowly, a theory of SISs can be a well
approximate tool to analyze the data sequences from them. However, there
are no asymptotic  limit sets in the SVSs. Consequently, it is inconvenient to
take the dynamics of the SVSs as chaotic dynamics although the real dynamics
still reveals sensitivity to initial conditions. Even in these SIS, transferred
from SVSs, there are still no chaotic attractors at all (will be discussed
later). An argument about the physical meaning for these calculated
 $D_2$ and $\lambda$ is the second  motivation of this Letter.

Let us begin with a typical developing digram [13] shown in Fig. 1 for the
logistic map
\begin{equation}
x_{n+1} = f(\mu, x_n) = 1 - \mu x_n^2,
\end{equation}
in the parameter interval $\mu \in [1.6, 1.8]$.
In the figure, we cover the parameter range $\mu \in [1.6, 1.8]$ by small
steps  $\delta \mu= 10^{-5}$ (as that usually used in bifurcation diagrams).
At $\mu = 1.6$, an initial value of $x_0$ is chosen which may be a point
on the attractor. We iterate the Eq. 1 only once with $\mu = 1.6$ and
obtain a point $x_1$. In the successive parameter $\mu = 1.6 + \delta \mu$,
an iterate $x_2$  of Eq. 1 is obtained starting from $x_1$ by considering the
continuously evolutionary process of  natural systems. We repeat it until
we reach $\mu = 1.8$ and draw all the pairs (1.6, $x_0$), (1.6 + $\delta \mu$,
 $x_1$), (1.6 + 2$\delta \mu$, $x_2$), $\cdots$ in a figure. In this way we get
 the $developing\ diagram$ in $x\ -\ \mu$ coordinates
as shown in Fig. 1. In the process, no point is thrown away
as transient. This developing diagram reflects the  evolutionary process of
a SVS modelled by the logistic map with a definite rapidity
 (the parameter step $\delta \mu$ reflects the rapidity of evolutionary
process of the dynamical system). It can be taken as the simplest
example of SVSs. In [13], the basic properties of this developing diagram
such as the discontinuity of period-doubling bifurcations, the time arrow [14]
and a brief comparison with those of the bifurcation diagram had been
discussed.

Now we test the short term predictability for the above developing diagram.
Considering that there is a data sequence $x_1,\ x_2,\ \cdots$ from a SIS,
according to Takens' theorem [4], there exists a function $F$ such that
\begin{equation}
x_n=F(x_{n-m\tau},x_{n-(m-1)\tau},\cdots,x_{n-\tau}),
\end{equation}
where $m$ is the embedding dimension, and $\tau$ is the time-delay.
The problem of predictability is how to find a good estimate of $F$ based on
the past history of $x_n$, on which various techniques of nonlinear
deterministic prediction have been developed.
In this Letter, we will use  the prediction algorithm [8] which is based on the
wavelet analysis and the neural networks [15]. This prediction
 algorithm  has succeeded in testing the time series from many dynamical models
such as the Ikeda map, the Lorenz equations, the Ushiki map, the
 Mackey-Glass differential-delay equations, etc for fixed parameters [8].

Suppose $\psi(x)~(x\in R^n)$ is a wavelet function, which can be thought of as
a band-pass function. Define
$$g(x)=\sum^N_{i=1} w_i\psi (D_iR_i(x-b_i))+\overline{g},\quad x\in R^n, $$
where the $b_i$'s are arbitrary translation vectors; the $D_i$'s are diagonal
 matrices built from arbitrary  dilation vectors (i.e., $D_i=$diag$(d_i)$,
 $d_i=(d_{1i},d_{2i},\cdots,d_{ni})$ is the dilation vector); the $R_i$'s are
 rotation matrices, which are used to compensate for the orientation selective
 nature of the dilations; the parameter $\overline{g}$ is introduced to help
 dealing with nonzero mean functions on finite domains (since
the wavelet $\psi(x)$ is zero mean); and the $w_i$ are weight coefficients.
It had been proved [15] that for any function in $L^2(R^n)$, there exist some
$g(x)$ with proper parameters such that the function can be approximated by
$g(x)$. Our  prediction algorithm [8] was proposed by determining  the free
parameters of $g(x)$ such that $g(x)$ can be a good approximation to the
function $F$ in Eq. 2. And the wavelet function $\psi(y)$ was chosen
explicitly as $$\psi(y)=\omega(y_1)\omega(y_2)\cdots \omega(y_m)$$ with [15]
$$	\omega(y_i)=(1-y_i^2)e^{-\frac{y_i^2}{2}},$$
where $y_i, i=1,2,\cdots,m$ are $m$ components of the reconstructed vectors.

For the data sequences $x_0, x_1, \cdots$ from the developing diagram
in Fig. 1, we fix the embedding dimension $m=2$ (will be discussed later),
and predict one step ahead. We take arbitrary 2000 data points in
succession from the developing diagram in Fig. 1. The first 500 points are
taken as input to determine the parameters in the prediction algorithm then
we use thenext 1500 points as the test points of prediction.
All of our numerical results are very satisfactory.
In Figs. 2-3 we show typical prediction results
 and the absolute prediction errors for the parameter range [1.74,1.76].
It is remarkable to find that the algorithm works so well that it can even
 predict the period 3 motion. Similar observations have been obtained
 for different selected $\delta \mu$ as well as in other range of parameters.
  We emphasize that the developing structure with variable
parameters of the logistic map (which is a SVS) including its periodic motion
and its period-doubling counterparts can be well predicted with a sophisticated
prediction algorithm for SISs.

In fact, the developing diagram shown in Fig. 1 is a trajectory in the phase
 portrait of the following two-dimensional (2D) map
\begin{equation}
\cases{x_{n+1} =  f(\mu_n, x_n) = 1 - \mu_n x_n^2,  \cr \mu_{n+1} =\mu_n +
\delta \mu,}
\end{equation}
from a 2D initial point $(x, \mu)$ = $(x_0, 1.6)$. The only
parameter in this 2D map is $\delta \mu$. Once the parameter $\delta \mu$
is fixed, the developing structure of the logistic map shown in Fig. 1 can be
described by a 2D  SIS (2).  As that has  been shown above, we do
have succeeded in predicting the developing structure of the logistic map shown
in Fig. 1 with a 2D prediction algorithm ($m$=2) for SISs.

Now we get the first conclusion for this Letter. A SVS can be discussed in its
variable-parameter space in which the SVS is transferred into a
higher-dimensional SIS so that the data sequence is short-term predictable.
We have checked this  idea for many other dynamical systems.  In Fig. 4 we
show our prediction results  for the   developing structure of the
H\'enon map\begin{equation}
\cases{x_{n+1} = 1 - a x_n^2 + y_n, \cr y_{n+1} = b x_n,}
\end{equation}
with parameters $(a,b)$ change simultaneously as following
\begin{equation}
\cases{a_{n+1} = a_n + a_n^3 \delta a, \cr b_{n+1} = b_n +  a_n \delta b,}
\end{equation}
where $a_0$ =1.52, $b_0$=0.1,  $\delta a$ = 2*10$^{-6}$ and $\delta b$ =
3*10$^{-6}$. Our prediction algorithm still works very well.

{}From the above discussion  we can understand the numerical result in [12] for
the Chinese national financial expenditure from 1953 to 1992 and other economic
data series. Since the natural environment
and the political structure had not changed much from 1953 to 1992, the
evolutionary process of the economical environment in this period might be
 governed by some equations similar to the Eq. 5 (though we might not
 know exactly what they are). In this variable-parameter space
the system is a SIS so that the data sequence might be well predicted. With
this idea, we have successfully tested some seismological data in China
recently [16].
Now we come to the second point in this Letter. How could a data sequence
from a SVS be characterized? So far, there are many sophisticated ways of
 characterizing
chaotic attractors, among which the Lyapunov exponents and the correlation
 dimensions are widely used. In the last decade, these techniques had been
 applied to the data sequences from many  natural systems. $D_2$
and $\lambda$ are calculated and their existence has been
considered as evidence of chaotic dynamics in those natural systems
[e.g.,9-11].
In fact, many of these natural systems are SVSs. Only when
the environment parameters change sufficiently slowly, these technique can be
approximately used to analyze the data  sequences from these natural systems.
A SVS does not exhibit chaotic attractor since there is no
asymptotic state. For these SISs from SVS in the variable-parameter space as
Eq. 3,  there is no unstable periodic orbit in their asymptotic
 state since the parameter $\mu$ always increases. As a result, it seems
inconvenient to  apply the ways of characterizing chaotic attractors to analyze
 the data sequences from SVSs although these data sequences might be
predictable.  However, $\lambda$ and $D_2$ did have been calculated for
the data sequences for some natural systems [e.g.,9-11] by using
the usual algorithms for chaotic attractors. What do these calculated
$\lambda$
and $D_2$ mean? Are they the evidence that the environment parameters for
these natural systems  change so slowly that they can be approximately taken as
SISs? Could the existence of those quantities be taken as evidence of chaotic
dynamics in those natural systems? In order to answer these questions we have
calculated the values of  $\lambda$ and $D_2$ for 10,000 data sequences for the
developing structure of the H\'enon map in Eqs. 4 and 5 with the algorithms
by [17,18]. They exist and  $D_2$=1.25$\pm$0.1 and  $\lambda$=0.294$\pm$0.002.
We also have computed these quantities for other developing parameter ranges of
the H\'enon map and the logistic map. From these results, it is clear that the
existence of the calculated $\lambda$ and  $D_2$ does not represent that the
system evolves so slowly that it can be
 approximated by a SIS. However, the positive value of $\lambda$
 are consistent with the sensitivity to initial conditions of these nonlinear
 dynamical systems in our calculations and a little smaller than the largest
 Lyapunov exponents for the dynamical systems with fixed parameters in the
 developing  parameter range  except for periodic orbits,
 and the correlation dimensions are greater than  those for the fixed
 parameters of which the above developing diagram sweeps out (For the H\'enon
map, $D_2$=1.03$\pm$0.02,  $\lambda$=0.353$\pm$0.003 for
$(a,b)$=(1.52,0.1) and  $D_2$=1.07$\pm$0.01, $\lambda$=0.321$\pm$0.001
for $(a,b)$=(1.5955,0.14671), corresponding to the first and the last datum,
 respectively). The physical meaning for these  calculated quantities
and how to characterize the data sequences from SVS still need to be
investigated, which will be presented in an extended paper.

 It should be noted that a data sequence from SVS can be easily confused with a
 time series of a chaotic attractor with noise. Fig. 5 shows the plotting  for
successive $x_{n+1}$ to $x_n$ for the same data sequence used above for
calculating $D_2$ and $\lambda$.
 This plotting looks very similar to what one would get for  a time series
 of a chaotic attractor with noise.  How to distinguish a data sequence
 between that from a SVS and from a chaotic attractor with or without noise
is still undertaken.
To conclude, we will say that the data sequences from some SVSs are predictable
with a nonlinear SIS prediction algorithm although the SVSs do not exhibit
chaotic attractors. This result encourages us to apply the prediction algorithm
to predict the data sequences from natural systems no matter their environment
parameters change slowly or rapidly. However, there are still some theoretical
problems needed to be clarified if one wants to apply the usual ways of
characterizing the chaotic attractors to  characterize the data sequences from
 SVSs otherwise $misleading$ results might be obtained. At last we noted that
 all the discussions in this Letter can be extended to discuss the dynamical
 systems described by differential or difference-differential  equations.

H.P.F thanks Prof. Hao Bailin for his encouragement.

\section*{}
{\bf Figure Captions}
\begin{description}
\item[]{ Fig. 1 \ The developing diagram of the logistic map in the parameter
interval $\mu \in [1.6,1.8]$ with $\delta \mu = 10^{-5}$.}
\item[]{ Fig. 2 \ One-step ahead prediction results for the developing
structure of the logistic map shown in Fig. 1 with $\mu \in [1.74,1.76]$.
 The crosses are the true data points and the
diamonds are the corresponding predictive ones. Only one point in each 20
data points is plotted.}
\item[]{ Fig. 3 \ Prediction error for Fig. 2. $x_T$ and $x_P$
are the true and the corresponding predictive values respectively.}
\item[]{ Fig. 4 \ Prediction error from the developing
structure of the H\'enon map with its parameters changes nonlinearly
shown in Eq. 5.}
\item[]{ Fig. 5 \ The plotting for successive $x_{n+1}$ to $x_n$ for the first
10,000 data from Eq. 4. }
\end{description}
\newpage
\begin{description}
\item[{*}]{ Present Address and address for correspondence:
Physics Department, Fudan University, Shanghai,  200433, P. R. Chinai;
email: zlin@fudan.ihep.ac.cn with subject "to Fang"}.
\item[{[1]}]{ E. Ott. Rev. mod. Phys. {\bf 53}, 655(1981)}.
\item[{[2]}]{ N.B. Abraham, J.P. Gollub, and H.L. Swinney, Physica {\bf 11}D,
252(1984)}.
\item[{[3]}]{ N. Packard, J. crutchfield, J.D. Farmer, and R. Shaw,
Phys. Rev. Lett.  {\bf 45}, 712(1980)}.
\item[{[4]}]{ F. Takens, in {\it Dynamical Systems and Turbulence}, edited by
D.A. Rand and L.-L. Young (springer-Verlag, Berlin, 1981);
 T. Sauer, J.A. Yorke, and M. Casdagli, J. Stat. Phys. {\bf 65}, 579(1991)}.
\item[{[5]}]{ P. Grassberger {\it et al.},
Chaos  {\bf 3}, 127(1993); E.J. Kostelich and T. Schreiber, Phys. Rev. E
{\bf 48}, 1752(1993)}.
\item[{[6]}]{ T. Shinbrot, C. Grebogi, E. Ott, and J.A. Yorke, Nature (London)
 {\bf 363}, 411 (1993)}.
\item[{[7]}]{ J.D. Farmer and J.J. Sidorowich, Phys. Rev. Lett. {\bf 59},
845(1987);M. Casdagli, Physica {bf 35}D, 335(1989);
H.D.I. Abarbanel, R. Brown and J.B. Kadtke, Phys. Rev. A {\bf 41}, 1782(1990);
J. Stark, J. Nonlinear Science, {\bf 3}, 197(1993)}.
\item[{[8]}]{ L.Y. Cao, Y.G. Hong, and H.P. Fang, submitted to Physica D.}
 \item[{[9]}]{C. Nicolis and G. Nicolis, Proc. Nat. Acad. Sci. USA, {\bf 83},
 536(1986).}
\item[{[10]}]{P. Chen, System Dynamics Review, {\bf 4}, 605(1988).}
\item[{[11]}]{G.P. Pavlos {\it et al.}, Int. J. Bif. \& Chaos, {\bf 4},
 87(1994).}\item[{[12]}]{ L.Y. Cao {\it et al.}, submitted to Comput.
Economics. In this
paper, several Chinese macroeconomic data sequences, e.g., the national
financial expenditure, total value of retain sales, gross output value of
industry and monetary supply, are found short-term predictable.}
\item[{[13]}]{ H.P. Fang, submitted to Phys. Rev. Lett.}
\item[{[14]}]{I. Prigogine, {\it From Being to Becoming},  San Francisco:
 W.H. freeman (1980)}.
\item[{[15]}]{Q.-H. Zhang and A. Benveniste, IEEE Trans. On. Neural Networks,
{\bf 3}, 889(1992);
C. Chui, {\it Wavelet: An introduction}, (New York, Academic, 1992)}.
\item[{[16]}]{ L.Y. Cao and H.P. Fang, Int. J. Bif. $\&$ Chaos (to be
published). In this paper, the $b$-value for the seismic events in
Tangshan region in China is well tested with our prediction algorithm.}
\item[{[17]}]{ J.-P. Eckmann {\it et al.},
Phys, Rev. A {\bf 34}, 4971 (1986)}.
\item[{[18]}]{ P. Grassberger and I. Procaccia,
Physica {\bf 9}D, 189 (1983)}.

\end{description}
\vfil

\end{document}